\documentclass{ifacconf}

\usepackage{graphicx}      % include this line if your document contains figures
\usepackage{natbib}        % required for bibliography
\usepackage{xcolor}
\usepackage{amsmath}
\usepackage{amssymb}
\usepackage{flushend}
\usepackage{algorithm}
\usepackage{algpseudocode}
\usepackage{enumitem}
\usepackage{algorithm}
\usepackage{epstopdf}
\usepackage{placeins}
\newtheorem{theorem}{Theorem}
\newtheorem{assumption}{Assumption}
\newtheorem{remark}{Remark}
\allowdisplaybreaks
\newcommand{\tr}{^{\intercal}} % define transpose easily
\newcommand\Tstrut{\rule{0pt}{2.6ex}}         % = `top' strut
\begin{document}
\begin{frontmatter}
	
	\title{Robust Adaptive Model Predictive Control with Worst-Case Cost} 
	% Title, preferably not more than 10 words.
	
	\thanks[footnoteinfo]{This work is supported by the Swiss National Science Foundation under grant number: 200021\_178890, and is part of the Swiss Competence Center for Energy Research SCCER FEEB\&D  of the Swiss Innovation Agency Innosuisse.}
	
	\author[First]{Anilkumar Parsi} 
	\author[First]{Andrea Iannelli} 
	\author[First]{Mingzhou Yin} 
	\author[First]{Mohammad Khosravi} 
	\author[First]{Roy S. Smith}
	
	\address[First]{Automatic Control Laboratory, ETH Zurich, Switzerland \\ (e-mail: \{aparsi,iannelli,myin,khosravm,rsmith\}@control.ee.ethz.ch).}
	
	\begin{abstract}                % Abstract of not more than 250 words.
		A robust adaptive model predictive control (MPC) algorithm is presented for linear, time invariant systems with unknown dynamics and subject to bounded measurement noise. The system is characterized by an impulse response model, which is assumed to lie within a bounded set called the feasible system set. Online set-membership identification is used to reduce uncertainty in the impulse response. In the MPC scheme, robust constraints are enforced to ensure constraint satisfaction for all the models in the feasible set. The performance objective is formulated as a worst-case cost with respect to the modeling uncertainties. That is, at each time step an optimization problem is solved in which the control input is optimized for the worst-case plant in the uncertainty set. The performance of the proposed algorithm is compared to an adaptive MPC algorithm from the literature using Monte-Carlo simulations. 
	\end{abstract}
	
	
\end{frontmatter}
	%===============================================================================

\section{Introduction}
% introduce adaptive control: attention
Model predictive control (MPC) is a popular strategy used to solve multivariable control problems due to its ability to handle nonlinearities and constraints while guaranteeing feasibility and stability (\cite{rawlings2009model}). The idea of MPC is to compute an input sequence at each time step, such that the input is optimal with respect to given performance index and system constraints. Only the first control input is applied to the plant, and the future control inputs are recomputed at the next time step. Robust MPC algorithms have the ability to handle uncertainty in the system models, ensuring that constraints are not violated despite inaccuracies in modeling the plant (\cite{bemporad1999robust}). However, having large model uncertainties adversely affects the performance of robust controllers. Adaptive control is one technique which can be used to ameliorate the conservatism induced by robustness to modeling inaccuracies. The idea of adaptive control is to carry out controller adjustments in real time based on the collected input-output data. Since MPC is an optimization based technique, performing online adaptation can be integrated easily into the MPC control structure. Utilizing this advantage, a variety of adaptive MPC control algorithms have been developed in the last decade. The main differences in these algorithms are in the model structure (impulse response, state-space, ARMAX, etc) and the adaptation techniques (set-membership identification, recursive least squares, etc) used. 

% difficult to do, so assumptions have to be made. Tell different kinds of work done over the years. Nonlinear systems, linear, state space, RLS estimates, ARMAX retc etc..
% Then you can start about specific to linear with FIR models, talk about how there was a bit of work done, and there are extensions.. 
In \cite{kim2008adaptive}, an adaptive MPC algorithm was developed for single input multiple output, linear time invariant (LTI) systems. The system was described using a state-space model with uncertain parameters which were identified online using a recursive least squares technique. In \cite{lorenzen2017} and \cite{lu2019}, adaptive MPC algorithms were proposed for multi input multi output (MIMO), linear time varying (LTV) systems, which were also described using a parametric state-space structure. The size of uncertainty in the model parameters was updated online using set-membership identification (\cite{milanese1991optimal}). Using results from robust tube MPC (\cite{kouvaritakis2016model}), these algorithms guarantee stability and recursive feasibility of the controllers. Instead of an estimate of the model, a worst-case cost was used to describe the performance of the MPC algorithm. This means that in addition to constraint satisfaction, the control performance is robust to the worst-case model uncertainty. However, the algorithms require the knowledge of the state-space structure and noise free measurements of the states, which can be restrictive. 

Alternatively, impulse response models have been used to describe the system dynamics. In \cite{tanaskovic2013}, an adaptive MPC algorithm was presented for single input single output (SISO), LTI systems with input and output constraints and measurement noise. The algorithm uses a finite impulse response (FIR) model, and assumes that the true impulse response of the system lies inside a bounded polytope. It was shown that the algorithm can handle large uncertainties, and hence the prior bounds on the FIR coefficients need not be tight. Set-membership identification was used to refine the model set online using measurement data. In \cite{tanaskovic2019adaptive}, the algorithm was extended to MIMO, LTV systems, with basis function parameterizations. In the algorithm, the MPC objective was defined using the Chebyshev center of the model set, while the constraints were robustly satisfied for all the models in the set. The control performance of the algorithm was improved in \cite{bujarbaruah2018adaptive} by using a recursive least squares estimator of the system, and defining chance constraints on the outputs. %A similar adaptive MPC algorithm was used for building climate control in \cite{tanaskovic2017robust}, demonstrating the practical applicability of the approach. The thermal model of the building was represented using basis functions, because the thermal dynamics in buildings are slow and require long impulse responses to be represented accurately.

In this paper, we present a robust adaptive MPC algorithm which uses an FIR model description and a worst-case performance index. For notational simplicity, SISO and LTI systems are considered in this paper but the results can be extended to MIMO and time varying systems. The system is subject to input constraints, output constraints and bounded measurement noise. To define the model uncertainty, the algorithm uses a polytopic feasible system set (FSS) which is updated online using set-membership identification. The constraints are enforced for all models in the FSS, and the objective function is defined using a min-max cost. Using such a cost function optimizes the control performance over all the plants in the FSS. The proposed controller guarantees recursive feasibility, and only requires the solution of linear and quadratic programs at each time step.  The performance of the robust adaptive MPC algorithm is compared against the adaptive MPC algorithm proposed in \cite{tanaskovic2013} to track different reference trajectories. It is shown that the robust adaptive MPC algorithm improves worst-case performance for all the trajectories considered, and the mean performance for some of the trajectories. 

\subsubsection{Notation:}
The sets of integers and real numbers are denoted by $ \mathbb{Z} $ and $ \mathbb{R} $ respectively, and the set of positive integers is denoted by $ \mathbb{Z}_{> 0} $. For a vector $ b $, $ b^{\intercal} $ represents its transpose, and $ [b]_i $ refers to the $ i ^{th}$ element in it. The $ i ^{th}$ row and $ j^{th} $ column of a matrix $ A $ are denoted by $ [A]_{i*} $ and $ [A]_{*j} $ respectively. The value $ x(i|t) $ denotes the value of the variable $ x $ at time step $ i $, predicted at the time step $ t $. The absolute value of a scalar $ a $ is denoted by $ |a| $. The rate of change of a signal  $ a(t) $ computed as $a(t)-a(t-1) $ is denoted by $ \Delta a(t) $. For any real scalar-valued function $ J $, $ \displaystyle\max_{h \in \mathbb{H}} J(h)$ refers to the maximum value of $ J $ over the set $ \mathbb{H} $.

\section{Background material}
\subsection{System description}
% assumptions
We consider a SISO, discrete time, strictly proper, LTI system. The system is assumed to have unknown but stable dynamics. The true system $S$ has an infinite impulse response (IIR) $ \{h_S\}_{1}^{\infty} $ where the elements $ \{h_S(1),h_S(2),\ldots\} $ are the impulse response coefficients of the system. To have a computationally tractable representation of the system for MPC, it is modeled using FIR coefficients.  They are represented by the components of the vector $ h_m \in \mathbb{R}^{m}$, where $ m $ is the length of the FIR. At any time step $ t\in \mathbb{Z}_{> 0} $, the system output $ y(t) $ can be represented as
\begin{equation}\label{eq:TrueOutput}
	y(t) = \displaystyle\sum_{i=1}^{\infty}h_S(i)u(t-i), %\doteq  {h_S* u(t)},
\end{equation}
and the model output $ y_m(t) $ as	
\begin{equation}\label{eq:ModelOutput}
y_m(t) = \displaystyle\sum_{i=1}^{m}h_m(i)u(t-i) \doteq {h_m* u(t)},
\end{equation}
where $ * $ is the convolution operator and $ {u(t)} $ is the input sequence.
The measured output of the system is 
\begin{equation*}
\tilde{y}(t) = y(t) + v(t), \quad \forall  t\in \mathbb{Z}_{> 0} ,
\end{equation*}
where $ v(t) $ is the measurement noise at the time step $ t $. The following assumptions are made on the system and the noise.
\begin{assumption}\label{As:noise}
	The noise $ v(t) $ is bounded according to
	\begin{equation}\label{eq:NoiseBound}
 |v(t)| < \epsilon, \qquad \forall t\in \mathbb{Z}_{> 0}.
	\end{equation}		
\end{assumption}
\begin{assumption}\label{As:hbound}
The IIR coefficients $ \{h_S\}_{1}^{\infty} $ satisfy the bounds
	\begin{align}
	\begin{split}\label{eq:hBounds}
L_l&\le h_S(i)\le L_u, \qquad i = 1,2,\ldots,\mu \\
L_l\rho^{i-\mu}&\le h_S(i)\le L_u\rho^{i-\mu},  i = \mu+1,\ldots,\infty,
	\end{split}
	\end{align}		
\end{assumption}
for parameters $ L_l,L_u,\rho \in \mathbb{R}$: $L_l,L_u \ge 0, \rho \in (0,1) $ and $ \mu \in \mathbb{Z}_{>0} $. 
\begin{remark}
Assumption \ref{As:noise} is reasonable because the measurement noise in most systems is bounded, and the bound is specified. Assumption \ref{As:hbound} is valid for open-loop stable systems (a common assumption in the context of system identification and adaptive control). The parameter $ \rho $ defines the rate of decay corresponding to the dominant pole in the system, while the parameters $ L_l$, $ L_u $ and $ \mu $ capture the initial dynamics. However, the assumption also restricts the sign of the IIR coefficients to be positive. This is because the algorithm presented here is suitable for systems which have uncertain parameters with a known sign. This is true for a certain class of systems, for e.g., positive systems (\cite{farina2011positive}), but the presented algorithm is not restricted to this class. It can be extended to LTI systems represented using basis functions (\cite{wahlberg1996approximation}), with the restriction that the signs of the basis function coefficients are known. 
\end{remark}

The system is subject to the input and output constraints given by
\begin{align}
\begin{split} \label{eq:Constraints}
|u(t)| \quad &\le \quad \bar{u}, \\
|\Delta u(t)|\quad  &\le \quad \overline{\Delta u}, \\
|y(t)| \quad  &\le \quad \bar{y},\qquad \forall t\in \mathbb{Z}_{> 0}.
\end{split}
\end{align}
The goal is to design a controller so that the output follows a known, desired trajectory $ y_\text{des} $ while satisfying the constraints \eqref{eq:Constraints}.

\subsection{Truncation error}	
% eta_m calculation: refer to [T2013]
The bounds on the FIR model coefficients $ h_m $ can be derived from Assumption \ref{As:hbound} as
	\begin{align}
\begin{split}\label{eq:hBoundsFIR}
L_l&\le h_m(i)\le L_u, \qquad i = 1,2,\ldots,\mu \\
L_l\rho^{i-\mu}&\le h_m(i)\le L_u\rho^{i-\mu},  i = \mu+1,\ldots,m.
\end{split}
\end{align}		
The bounds used in \eqref{eq:hBoundsFIR} are equal to the bounds on the first $ m $ IIR coefficients in \eqref{eq:hBounds}. However, the bounds in \eqref{eq:hBoundsFIR} can be relaxed since the proposed controller is adaptive. That is, the exact knowledge of the true system dynamics is not necessary, and an upper bound on the decay of its impulse response is sufficient for this algorithm. Using \eqref{eq:hBoundsFIR}, the model can account for the part of the output due to the first $ m $ impulse response coefficients of the true system. However truncating the length of the impulse response results in an error in the model's prediction, which can be bounded as
\begin{align}\label{eq:eta_m}
|y(t)-y_m(t)| &= 
\left|\displaystyle\sum_{i=m+1}^{\infty}h_S(i)u(t-i) \right| \nonumber \\ &\le \displaystyle\sum_{i=m+1}^{\infty}|h_S(i)u(t-i)| \nonumber \\ &\le
\bar{u}\displaystyle\sum_{i=m+1}^{\infty}|h_S(i)| \nonumber \\ &\le \bar{u} L \rho^{m-\mu}\frac{\rho}{1-\rho} \quad \doteq \: \eta_m. 
\end{align}

\subsection{Online set-membership identification}
Set-membership identification is a technique used to identify systems affected by noise with unknown statistical properties (\cite{milanese1991optimal}). Here, an initial FSS is defined as the set of all possible models that are consistent with the initial information
\begin{equation}\label{eq:H0}
H(0) := \left\{h_m \in \mathbb{R}^{m}\: | \eqref{eq:hBoundsFIR} \right\},
\end{equation}
which is a polytope in $ \mathbb{R}^{m}$. At each time step $ t $, a non\nobreakdash-falsified set is used to update the set $ H(t) $, which is the FSS at that time step. For systems with FIR descriptions, polytopic non\nobreakdash-falsified sets can be constructed from measurement data. For example, using the measurement $ \tilde{y}(t) $, a simple non\nobreakdash-falsified set can be written as
\begin{align} \label{eq:newCons}
\begin{split}
\delta(t) &:= \left\{h\in \mathbb{R}^{m} \bigr| \: |\tilde{y}(t) - h*u(t)| \le \eta_m + \epsilon \right\} \\
&= \left\{h \in \mathbb{R}^{m}\: \Biggr|\begin{array}{ll}
 h*u(t) &\le \tilde{y}(t) + \eta_m + \epsilon, \\
 h*u(t) &\ge \tilde{y}(t) - \eta_m - \epsilon \\
\end{array} \right\}\\
\end{split}
\end{align}
where $ \eta_m$ and $ \epsilon $ are defined according to \eqref{eq:NoiseBound} and \eqref{eq:eta_m} respectively and $ \delta(t) $ contains the set of all models that could have generated the measurement $ \tilde{y}(t) $. As proposed in \cite{chisci1998}, using a block of  $ s $ measurements $[\tilde{y}(t-s),\ldots,\tilde{y}(t-1)]$ to generate the non\nobreakdash-falsified set improves the identification. Let $ \phi(k) \in \mathbb{R}^{m} $ denote the vector of past $ m $ inputs at time step $ k $ arranged as
\begin{equation*} %\label{eq:Regressor}
\phi(k) = [u(k),u(k-1),\ldots,u(k-m+1)]\tr ,
\end{equation*}
then the convolution $ h*u(k) $ can be written as $ \phi(k)\tr h $. Similar to  \eqref{eq:newCons}, the following non\nobreakdash-falsified set is defined 
\begin{align}\label{eq:Delta_s}
\Delta_s(t) &:= \left\{h \in \mathbb{R}^{m}\: \Biggr|\begin{array}{ll}
\phi(k)\tr h &\le \tilde{y}(k) + \eta_m + \epsilon,  \vspace{0.5em} \\
\phi(k)\tr h &\ge \tilde{y}(k) - \eta_m - \epsilon, \\
& \forall k \in [t-1,t-s]
\end{array} \right\}\\
&= \left\{h \in \mathbb{R}^{m}\: | A_{\Delta}(t)h \le b_{\Delta} \right\}.\nonumber
\end{align}
where the matrices $ A_{\Delta} \in \mathbb{R}^{2s\times m}, b_{\Delta} \in \mathbb{R}^{2s} $ are used to characterize $ \Delta_s(t) $, which is the set of all models that could have generated the  measurements $[\tilde{y}(t-s),\ldots,\tilde{y}(t-1)]$. The FSS can be updated at each time step using $ \Delta_s(t) $ according to
\begin{equation}\label{eq:HRecUpdate}
 H(t) = H(t-1) \cap \Delta_s(t).
\end{equation}
Since the initial FSS is defined as a polytopic set and $ \Delta_s(t) $ is polytopic, the set $ H(t) $ remains polytopic $ \forall t $ if it is updated according to \eqref{eq:HRecUpdate}. However, the number of hyperplanes in $ H(t) $ will increase at every time step. To prevent this, $ H(t) $ is defined using a finite number of polytopic constraints given in
\begin{equation} \label{eq:HDef}
{H}(t) :=  \{h| A_h h\le b_h(t)\},
\end{equation}
where $ A_h \in \mathbb{R}^{p\times m}$ is a matrix chosen offline and $ b_h(t) \in \mathbb{R}^{p}$ is updated online such that $ H(t) \supseteq H(t-1) \cap \Delta_s(t) $ is satisfied. This is ensured by calculating $ b_h(t) $ as a solution to the following set of $ p $ linear programs:
\begin{align}\label{eq:Htupdate}
\begin{split}
[b_h(t)]_i \: =\: &\max_{h\in \mathbb{R}^{m}} \quad  [A_h]_{i*} h\\
 &\text{s. t. } \quad  \begin{bmatrix}
					 A_h \\ A_{\Delta}
					 \end{bmatrix} h
 \le \begin{bmatrix}
 b_h(t-1) \\ b_{\Delta}
 \end{bmatrix}, \quad i = 1,2,\ldots,p .\\
 \end{split}
\end{align}
\section{Robust adaptive model predictive control}
In MPC for systems with FIR models, an optimization problem is solved at each time step to calculate the control inputs for a finite prediction horizon. The control input at the end of the prediction horizon is chosen such that it can remain at a constant value after the horizon while satisfying constraints. This ensures recursive feasibility of the algorithm, whereby the optimization problem at the next time step remains feasible. The control input corresponding to the first time step is applied to the system, and the future control inputs are recalculated at the next time step. In this section, an optimization problem consisting of the worst-case cost with respect to model uncertainties is formulated, while enforcing the system constraints defined in \eqref{eq:Constraints}. A robust adaptive MPC algorithm is then described.

Let $ U $ be the vector of predicted control inputs $ [u(t|t),u(t+1|t),\ldots,u(t+N-1|t)] $, where $ N $ is the prediction horizon. Let $\{\tilde{u}(t-m +1),\tilde{u}(t-2),\ldots,\tilde{u}(t-1)\} $ be the previous $ m -1$ control inputs applied to the system. The vectors $ \phi(i|t) \in \mathbb{R}^{m} $ are constructed using past and predicted future control inputs as
\begin{align} \label{eq:phiDef1}
	\phi(i|t) &= \begin{bmatrix}
q(i|t) & \ldots &	q(i-m+1|t) 
	\end{bmatrix}\tr, \\
				& \qquad  i=t,t+1,\ldots,t+N+m-2, \nonumber
\end{align}
where $ q(k|t) $ is defined as
\begin{align}\label{eq:phiDef2}
q(k|t) = \left\{ \begin{array}{l l}
\tilde{u}(k) & \text{for } k<t \\
u(k|t) & \text{for } t<k<t+N-1 \\
u(t+N-1|t) & \text{for } k>t+N-1 .
\end{array} \right.
\end{align}	
The vectors $ \phi(i|t) $ are a combination of the past control inputs which are known, and the future control inputs which are the decision variables. These vectors will be used to represent the future outputs of the system $ y(i|t) $ as $ \phi(i|t)\tr h $ , for $ h\in H(t)$ and $ i \in [t,t+N+m-2] $.  In \eqref{eq:phiDef2}, the control inputs after the prediction horizon are set to be equal to $ u(t+N-1|t) $, so that recursive feasibility can be guaranteed.
%\begin{remark}
%	Note that though the prediction horizon is $ N $, $ N+m-1 $ future inputs are considered in the definition of the vectors $ \phi(i|t) $. The last $ m $ inputs are constrained to be equal, which essentially enforces a terminal constraint and ensures recursive feasibility of the algorithm. 
%\end{remark}

\subsection{Robust objective function}	
The optimization problem in the MPC algorithm includes a cost minimization objective. To ensure that the performance of the controller is robust to the worst-case plant uncertainty, the cost function $ J $ is defined according to
\begin{equation}\label{eq:Cost1}
	J = \max_{h \in H(t)} \displaystyle\sum_{i=t}^{t+N-1} \bigr(y_\text{des}(i|t) - \phi(i|t)\tr h\bigr)^{2},
\end{equation}
where $ y_\text{des} $ is the desired output trajectory and the vectors $ \phi(i|t) $ are defined in \eqref{eq:phiDef1} and \eqref{eq:phiDef2}. Additional terms which penalize the input $ u $ and the rate of change of input $ \Delta u $ can be added to the cost function, but are omitted for notational simplicity. The defined cost function ensures that the optimized control inputs minimize the maximum possible value of $ J $ over all the models in $ H(t) $. This is a min-max objective, which must be reformulated in a convex manner. Using supplementary cost variables $ c \in \mathbb{R}^{N} $, the cost function in \eqref{eq:Cost1} can be rewritten as
\begin{align}\label{eq:Cost2}
J = \displaystyle\sum_{j=1}^{N}& c(j)^{2}, \\
\begin{split}\label{eq:RobustCostCons}
\text{s.t }\quad \max_{h \in H(t)} \quad y_\text{des}(i|t) - \phi(i|t)\tr h &\le c(i-t+1),\\
\max_{h \in H(t)} \: -y_\text{des}(i|t) +\phi(i|t)\tr h &\le c(i-t+1),
\end{split}\\
& \forall i \in [t,t+N-1].\nonumber
\end{align}
In \eqref{eq:RobustCostCons}, each element of $ c $ is specified as an upper bound on the worst case deviation from the reference. 
\subsection{System constraints and optimization problem}
While the constraints \eqref{eq:RobustCostCons} are used to implement a min-max objective, the system's input and output constraints given in \eqref{eq:Constraints} must be enforced through the optimization problem. The constraints on the input at each time step can be written as
\begin{equation}\label{eq:inpCons}
\begin{array}{r l l}
 -\bar{u} \le u(i|t) &\le \bar{u}, \\
-\overline{\Delta u} \le \Delta u(i|t) &\le \overline{\Delta u}, \quad \forall i \in [t,t+N-1].
\end{array}
\end{equation}
The output constraints must be satisfied for all the models in $ H(t) $, i.e., robustly satisfied. The system outputs can be bounded according to 
\begin{align*}
y(i|t) \ge  \min_{h \in H(t)} \phi(i|t)\tr h  &- \eta_m, \\
y(i|t) \le  \max_{h \in H(t)} \phi(i|t)\tr h  &+ \eta_m, \\
 &\forall i \in [t+1,t+N+m-2],
\end{align*}
where $ \eta_m $ is the truncation error defined in \eqref{eq:eta_m}. Hence, the output constraints in \eqref{eq:Constraints} can be formulated as
\begin{equation}\label{eq:OutputCons}
\begin{array}{l r l }
\max_{h \in H(t)} &  \phi(i|t)\tr h &\le \bar{y} -\eta_m, \\
 \max_{h \in H(t)} & -\phi(i|t)\tr h &\le \bar{y}-\eta_m, \\
 & &\forall i \in [t+1,t+N+m-2], 
\end{array}
\end{equation}
where the outputs constraints for $ i \in [t+N,t+N+m-2] $ are enforced to ensure no constraint violations occur when a constant input is applied at the end of the prediction horizon. Thus, the optimization problem in the MPC controller can be set up as
\begin{equation}\label{eq:OptProb}
\begin{array} {r l}
\min_{U} & J \\
\text{subject to} & \eqref{eq:RobustCostCons}, \eqref{eq:inpCons},\eqref{eq:OutputCons}.
\end{array}
\end{equation}
The constraints \eqref{eq:RobustCostCons} and \eqref{eq:OutputCons} can be replaced by an equivalent set of linear constraints using techniques from robust optimization (\cite{ben2009robust}), as shown in Appendix \ref{sec:RobCons}. Using this reformulation, \eqref{eq:OptProb} is simplified to a standard quadratic program. 
\subsection{Robust adaptive MPC algorithm}
The procedure for robust adaptive MPC is described in Algorithm \ref{Alg:RAMPC}.
\begin{algorithm} [h]
	\caption{Robust adaptive MPC}\label{Alg:RAMPC} 
	\begin{algorithmic}[1]
		\Statex \textbf{Initialize} $ \Delta_s(0) $ using past measurements and inputs according to \eqref{eq:Delta_s}.
		\Statex \textbf{Initialize} H(0) according to \eqref{eq:H0},\eqref{eq:HDef}.
		\State $ t\gets 1 $
		\Repeat 
		\State Obtain the measurement $ \tilde{y}(t) $
		\State Update $ \Delta_s(t) $ according to \eqref{eq:Delta_s}.
		\State Update $ H(t) $ using \eqref{eq:Htupdate}.
		\State Solve optimization problem \eqref{eq:OptProb} to compute $ U $.
		\State Apply the control input  $ u(t|t) \gets U(1) $.
		\State $ t \gets t+1 $
		\Until 
	\end{algorithmic}
\end{algorithm}
%\begin{algorithm}[t]
%\caption{Robust Adaptive MPC}
%\label{Alg:RAMPC}
%\begin{enumerate}[label=\textbf{Step \arabic*},leftmargin=*,font=\small]
%\item At each time step t, update the feasible parameter set $ H(t) $ according to \eqref{eq:Htupdate}.
%\item Solve the optimization problem \eqref{eq:OptProb}.
%\item Apply the control input $ u(t) = u*(t|t) $, set t to t+1, go to \textbf{\small Step 1}.
%\end{enumerate}
%\end{algorithm}
The algorithm must be initialized with the set $ H(0) $ and the past inputs applied to the system. At each time step $ t $, the measurements $\tilde{y}$(t) are used to update the non-falsified set $ \Delta_s(t) $ and $ H(t) $ according to \eqref{eq:Delta_s} and \eqref{eq:Htupdate} respectively. Algorithm \ref{Alg:RAMPC} guarantees robust satisfaction of input and output constraints, as shown by the following theorem from \cite{tanaskovic2013}.
% Theorem: recursive feas
\begin{theorem}
	If Assumptions 1-2 hold, and the optimization problem \eqref{eq:OptProb} is feasible at t=0. Then, the closed loop system obtained by applying Algorithm \ref{Alg:RAMPC} is guaranteed to satisfy the input and output constraints $ \forall t>0 $.
\end{theorem}
% The proof relies on the fact that for all the models in the FSS at a time step, a feasible solution exists for all future time steps if the optimization problem is solved. Adapting the FSS 
In Algorithm \ref{Alg:RAMPC}, at each time step the FSS is truncated using measurement data. The performance of the closed loop system improves with time because the size of the uncertainty set decreases and hence the controller is adaptive. The advantage of including a worst-case cost in the objective function of the MPC controller is that it increases the robustness of the performance to model uncertainty. However, this affects its adaptivity, i.e., the amount of information it learns from a system. This is because the control inputs of robust adaptive controllers are generally smaller than that of a controller using nominal cost function. To investigate the effects of adaptivity and robustness on the overall performance on the system, a simulation study was performed.

\section{Simulations}
In this section, the performance of the robust adaptive MPC (RAMPC) algorithm proposed in Algorithm \ref{Alg:RAMPC} is compared against the adaptive MPC (AMPC) algorithm given in \cite{tanaskovic2013}. The AMPC algorithm defines the objective function of the MPC controller using the Chebyshev center of the FSS as the estimate of the true system. The objective function is defined therein as a quadratic function of the predicted 
deviation from the reference trajectory, along with penalties on the input and the rate of change of input. To make the comparison equivalent, the penalties on input and input rate have been removed. Thus in both the AMPC and RAMPC algorithms, the only objective of the optimal control problem is reference tracking. The performance of the algorithms is compared based on the root mean squared (RMS) deviation from the reference trajectory.

\subsection{Parameter Settings}
The true systems which must be controlled by the AMPC and RAMPC algorithms were randomly sampled from a predefined and bounded set. The parameters defining the initial FSS are given in Table \ref{tab:SysPar}, where the bounds on the impulse response coefficients can be calculated from \eqref{eq:hBounds}.
\begin{table}[h]
	\centering 
	\caption{Parameters describing bounds on impulse response, input and output}
	\label{tab:SysPar}
\begin{tabular}{llllllll}
	\hline 
	$ L_l $ & $ L_u $ & $ \mu $ & $ \rho $ & $ \epsilon $ & $\bar{u}$ & $ \overline{\Delta u}$ & $\bar{y}$ \Tstrut  \\
	\hline
	0.3 & 1 & 4 & 0.65 & 0.1 & 2 & 0.8 & 4\Tstrut  \\
	\hline 
\end{tabular} % L, eps, rho, mu, length..
\end{table}
%\begin{remark}
%	The parameters in table \ref{tab:SysPar} represent simplistic systems with quickly decaying impulse responses. This is because the parameters were chosen such that the Monte-Carlo simulations could be run in reasonable time. Systems with longer impulse responses can be chosen when controlling a real system, subject to the constraints on computation time. 
%\end{remark}

A set of 200 impulse responses was obtained by random sampling of each coefficient from a uniform distribution between the bounds defined in \eqref{eq:hBounds}. Each impulse response obtained was considered as a true system which must be controlled using the AMPC and RAMPC algorithms. The prediction horizon was set to $ N= $ 15 time steps, and the length of the FIR in the model was truncated to $ m = 12 $ parameters. This results in a truncation error $ \eta_m = 0.12$, as given in \eqref{eq:eta_m}. To define the initial FSS according to \eqref{eq:HDef}, a matrix $ A_h $ was constructed with $ p=156 $ constraints. The initial bounds $ b_h(0) $ were calculated using the parameters specified in Table \ref{tab:SysPar}. The value of $ b_h(t) $ was updated at each time step using a block of $ s = 36 $ past measurements. 
%The parameters for the controllers are given table \ref{tab:ContPar}.
%\begin{table}[h]
%	\centering
%	\begin{tabular}{llll}
%		\hline
%		$ N $ & $ m $ & $ p $ & $ s $   \\
%		\hline
%		10 & 7 & 56 & 7  \\
%		\hline 
%	\end{tabular} % N, m, p, s
%	\caption{Controller parameters for adaptive MPC}
%	\label{tab:ContPar}
%\end{table} 

\begin{figure}
	\includegraphics{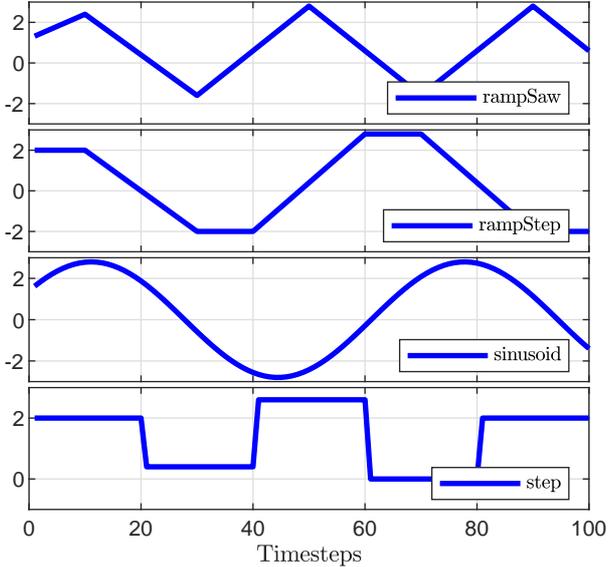}
	\centering
	\caption{Reference trajectories used in Monte-Carlo simulations to compare the performance of nominal and robust adaptive MPC.}
	\label{fig:RefTraj}
\end{figure}

In both the algorithms, the control inputs form regressor vectors which are used to refine the FSS as shown in \eqref{eq:Delta_s} and \eqref{eq:Htupdate}. The reference trajectory to be tracked influences the sequence of control inputs, and hence the refinement of the uncertainty sets. However, only a step reference was considered in \cite{tanaskovic2013}. To investigate dependence of controller performance on the reference trajectory, AMPC and RAMPC algorithms were used to track the different trajectories in Figure \ref{fig:RefTraj}. The step trajectory has a quick change in reference which is difficult to track. A smaller peak to peak amplitude was chosen for the step so that the overall deviations from different trajectories have comparable magnitudes.
% The maximum value to be tracked in each of these trajectories is in the range [2.6 2.7]. 
The optimization problems in both the algorithms were solved using CVX (\cite{cvx}, \cite{gb08}) and  MOSEK (\cite{andersen2000mosek}). 
\begin{figure*}[!ht]
	\includegraphics[width=\textwidth]{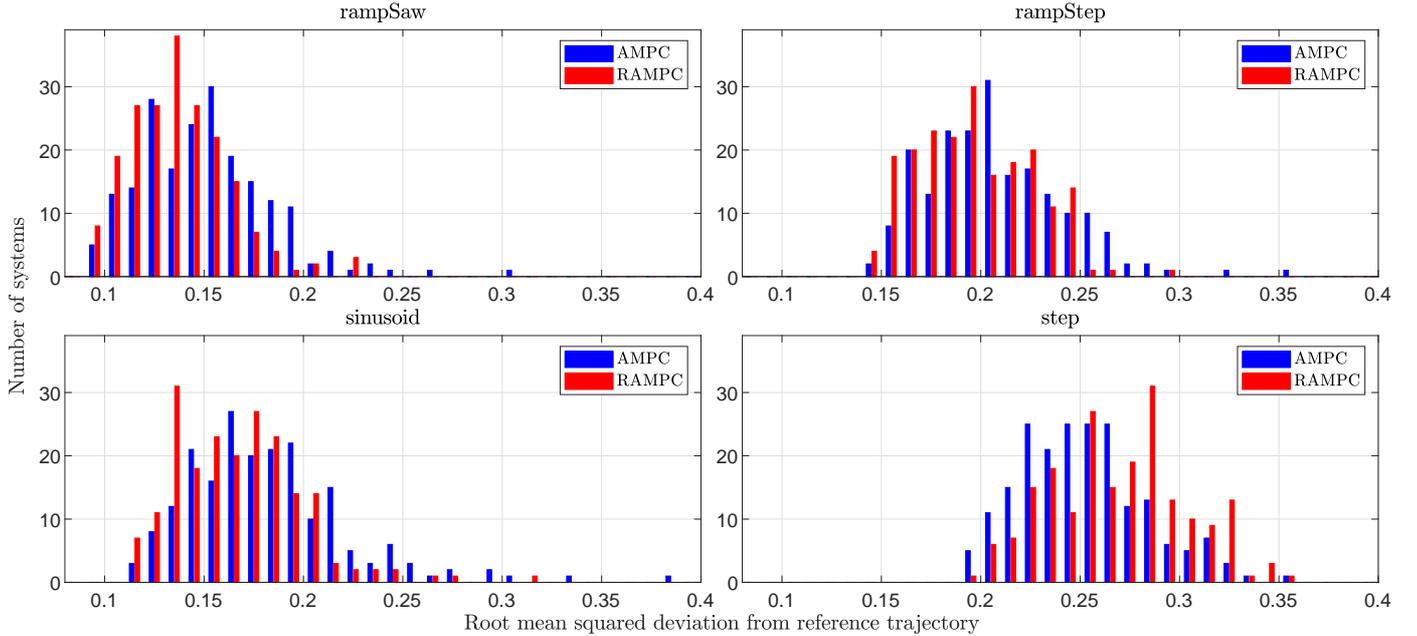}
	\caption{Distribution of RMS deviation of output from each reference trajectory using AMPC and RAMPC algorithms.}
	\label{fig:dist}
\end{figure*}
\begin{figure}[t]
\includegraphics{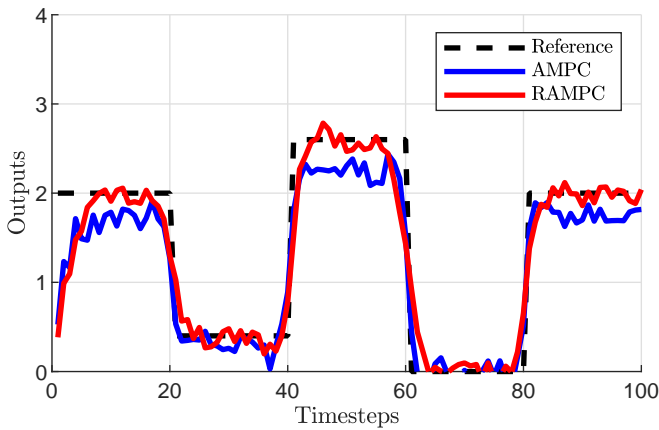}
\centering
\caption{Worst-case tracking performance of the AMPC and RAMPC algorithms using a step reference.}
\label{fig:WorstCaseStep}
\end{figure}
\subsection{Results and discussions}
The performance of the algorithms was evaluated based on the RMS deviation from the given reference trajectory. Figure \ref{fig:dist} shows the distribution of the RMS deviations from each of the trajectories considered. The histograms in blue and red represent the performance of the AMPC and RAMPC algorithms respectively. Since the FSS is in a high dimensional space ($ m=12 $), its center might not always be a good representation of the true system. For all trajectories except the step, using a robust controller improves the control performance even when the worst-case uncertainty is not realized. It can be seen that the distribution of RMS deviation for the RAMPC algorithm has a smaller tail on the right side. This corresponds to better performance using the RAMPC algorithm when the true system is not close to the Chebyshev center of the FSS.  In addition, the performance of the algorithms is dependent on the reference trajectory, as seen in the plot corresponding to the step reference. Due to the quick changes in the trajectory, the minimum value of RMS deviation measured for the step reference was higher. %However, the variance of the RMS deviation is the least for the step reference. This is because quick changes in the input signals were applied to track the step reference,  which improved the refinement of the FSS. Hence, the step trajectory has a better worst-case performance and a lower variance in the total deviation. 

The mean and maximum of the deviations from reference trajectories are given in Table \ref{tab:MedMaxData} and the results with better performance are highlighted. The maximum of the RMS deviation corresponds to the worst-case performance, which is improved for all the trajectories while using the RAMPC algorithm. Comparing the means of the RMS deviation from reference, it can be seen that RAMPC performs better for all the trajectories except the step. This is due to the robust cost in the RAMPC algorithm, which results in conservative responses while tracking the quick changes in the step reference. 
\begin{table}[h!]
	\centering
	\caption{Mean and maximum values of the root mean squared deviations from each trajectory, tracked using AMPC and RAMPC algorithms.}
	\label{tab:MedMaxData}
	\begin{tabular}{|c|c|c|c|c|}
		\hline
	&\multicolumn{2}{|c|}{Mean RMS deviation}&		\multicolumn{2}{|c|}{Maximum  RMS deviation}  \Tstrut\\
		\cline{2-5}
		Trajectory	&AMPC & RAMPC & AMPC & RAMPC \Tstrut \\
		\hline
\Tstrut rampSaw  & 0.152 & \textbf{0.137}  & 0.303  &  \textbf{0.228} \\
		rampStep & 0.208 & \textbf{0.196} & 0.352  &  \textbf{0.292} \\
		sinusoid & 0.182 & \textbf{0.167}  & 0.380  &  \textbf{0.311} \\
		step     & \textbf{0.251} & 0.269 & 0.354  &  \textbf{0.350} \\

		\hline 
	\end{tabular} % L, eps, rho, mu, length..
	\vspace{0.1cm}
\end{table}

In Figure \ref{fig:WorstCaseStep}, the trajectories with the highest RMS deviation from the step reference are plotted, i.e., the trajectories from the worst-case plant. It can be seen that the tracking performance improves with time due to the reduction in the model uncertainty. The conservatism of the RAMPC algorithm while tracking the quick changes is visible in the trajectories. However, the flat portion of the trajectory is tracked better with the RAMPC algorithm, and it reduces the worst-case cost of the AMPC algorithm. The computational complexity of the optimal control problem is higher in the RAMPC algorithm due to the additional robust constraints defined in \eqref{eq:RobustCostCons}. However, the AMPC algorithm estimates the Chebyshev center at each time step by solving an additonal linear program. Table \ref{tab:Runtime} compares the runtimes of the algorithms at each time step, when the number of FIR coefficients was incrementally increased from 8 to 20.
\begin{table}[ht!]
	\centering
	\caption{Runtimes for different FIR lengths}
	\label{tab:Runtime}
	\begin{tabular}{c c c}
		\hline
		$ m $ & $ t_\text{AMPC} $[in s] & $ t_\text{RAMPC} $[in s]    \\
		\hline
		8 & 2.10 & 1.80  \\
		10 & 2.37 & 2.32 \\
		12 & 2.44 & 2.52  \\
		14 & 2.92 & 3.10 \\
		20 & 7.82 & 8.80 \\
		\hline 
	\end{tabular} % N, m, p, s
\end{table} 

 The values $ t_\text{AMPC} $ and $ t_\text{RAMPC} $ correspond to the time required to calculate new control inputs at each time step for the AMPC and RAMPC algorithms respectively. The simulations were performed on an Intel i7-8550U 1.8 GHz processor, by keeping the prediction horizon constant. It can be seen that for small impulse responses, the RAMPC algorithm is faster. The runtime of the RAMPC algorithm is higher for larger systems, but is only 12\% higher than the corresponding runtime for the AMPC algorithm.

% histogram results
% max cost
% average cost
% median performance results
% worst case error results
\section{Conclusion}
In this paper, a robust adaptive MPC algorithm was described which can be applied for LTI systems with input and output constraints. The algorithm uses set online membership identification to reduce the uncertainty in the model parameters, and ensures robust constraint satisfaction. The objective function of the MPC algorithm is designed so that it is robust to model uncertainty, and the resulting optimization problem was posed as a standard quadratic program. The algorithm was compared to an existing algorithm in literature using Monte Carlo simulations. The simulation study showed that using a worst-case objective function improves the performance of the adaptive MPC controller for various reference trajectories. Future work will focus on extending the algorithm to a more general class of systems using a basis function representation.

\bibliography{adaptiveMPC}             % bib file to produce the bibliography

\begin{thebibliography}{18}
\providecommand{\natexlab}[1]{#1}
\providecommand{\url}[1]{\texttt{#1}}
\providecommand{\urlprefix}{URL }
\expandafter\ifx\csname urlstyle\endcsname\relax
  \providecommand{\doi}[1]{doi:\discretionary{}{}{}#1}\else
  \providecommand{\doi}{doi:\discretionary{}{}{}\begingroup
  \urlstyle{rm}\Url}\fi

\bibitem[{Andersen and Andersen(2000)}]{andersen2000mosek}
Andersen, E.D. and Andersen, K.D. (2000).
\newblock The {MOSEK} interior point optimizer for linear programming: an
  implementation of the homogeneous algorithm.
\newblock In \emph{High Performance Optimization}, 197--232. Springer.

\bibitem[{Bemporad and Morari(1999)}]{bemporad1999robust}
Bemporad, A. and Morari, M. (1999).
\newblock Robust model predictive control: A survey.
\newblock In \emph{Robustness in Identification and Control}, 207--226.
  Springer.

\bibitem[{Ben-Tal et~al.(2009)Ben-Tal, El~Ghaoui, and
  Nemirovski}]{ben2009robust}
Ben-Tal, A., El~Ghaoui, L., and Nemirovski, A. (2009).
\newblock \emph{Robust Optimization}, volume~28.
\newblock Princeton University Press.

\bibitem[{Bertsimas and Tsitsiklis(1997)}]{bertsimas1997introduction}
Bertsimas, D. and Tsitsiklis, J.N. (1997).
\newblock \emph{Introduction to linear optimization}, volume~6.
\newblock Athena Scientific Belmont, MA.

\bibitem[{Bujarbaruah et~al.(2018)Bujarbaruah, Zhang, and
  Borrelli}]{bujarbaruah2018adaptive}
Bujarbaruah, M., Zhang, X., and Borrelli, F. (2018).
\newblock Adaptive {MPC} with chance constraints for {FIR} systems.
\newblock In \emph{2018 Annual American Control Conference (ACC)}, 2312--2317.

\bibitem[{Chisci et~al.(1998)Chisci, Garulli, Vicino, and Zappa}]{chisci1998}
Chisci, L., Garulli, A., Vicino, A., and Zappa, G. (1998).
\newblock Block recursive parallelotopic bounding in set membership
  identification.
\newblock \emph{Automatica}, 34(1), 15--22.

\bibitem[{Farina and Rinaldi(2011)}]{farina2011positive}
Farina, L. and Rinaldi, S. (2011).
\newblock \emph{Positive Linear Systems: Theory and Applications}, volume~50.
\newblock John Wiley \& Sons.

\bibitem[{Grant and Boyd(2008)}]{gb08}
Grant, M. and Boyd, S. (2008).
\newblock Graph implementations for nonsmooth convex programs.
\newblock In V.~Blondel, S.~Boyd, and H.~Kimura (eds.), \emph{Recent Advances
  in Learning and Control}, Lecture Notes in Control and Information Sciences,
  95--110. Springer-Verlag Limited.

\bibitem[{Grant and Boyd(2014)}]{cvx}
Grant, M. and Boyd, S. (2014).
\newblock {CVX}: Matlab software for disciplined convex programming, version
  2.1.
\newblock \url{http://cvxr.com/cvx}.

\bibitem[{Kim and Sugie(2008)}]{kim2008adaptive}
Kim, T.H. and Sugie, T. (2008).
\newblock Adaptive receding horizon predictive control for constrained
  discrete-time linear systems with parameter uncertainties.
\newblock \emph{International Journal of Control}, 81(1), 62--73.

\bibitem[{Kouvaritakis and Cannon(2016)}]{kouvaritakis2016model}
Kouvaritakis, B. and Cannon, M. (2016).
\newblock \emph{Model Predictive Control: Classical, Robust and Stochastic}.
\newblock Springer, London.

\bibitem[{Lorenzen et~al.(2017)Lorenzen, Allg{\"o}wer, and
  Cannon}]{lorenzen2017}
Lorenzen, M., Allg{\"o}wer, F., and Cannon, M. (2017).
\newblock Adaptive model predictive control with robust constraint
  satisfaction.
\newblock \emph{IFAC-PapersOnLine}, 50(1), 3313--3318.

\bibitem[{Lu and Cannon(2019)}]{lu2019}
Lu, X. and Cannon, M. (2019).
\newblock Robust adaptive tube model predictive control.
\newblock In \emph{2019 American Control Conference (ACC)}, 3695--3701.

\bibitem[{Milanese and Vicino(1991)}]{milanese1991optimal}
Milanese, M. and Vicino, A. (1991).
\newblock Optimal estimation theory for dynamic systems with set membership
  uncertainty: an overview.
\newblock \emph{Automatica}, 27(6), 997--1009.

\bibitem[{Rawlings and Mayne(2009)}]{rawlings2009model}
Rawlings, J.B. and Mayne, D.Q. (2009).
\newblock \emph{Model predictive control: Theory and design}.
\newblock Nob Hill Pub.

\bibitem[{Tanaskovic et~al.(2019)Tanaskovic, Fagiano, and
  Gligorovski}]{tanaskovic2019adaptive}
Tanaskovic, M., Fagiano, L., and Gligorovski, V. (2019).
\newblock Adaptive model predictive control for linear time varying {MIMO}
  systems.
\newblock \emph{Automatica}, 105, 237--245.

\bibitem[{Tanaskovic et~al.(2013)Tanaskovic, Fagiano, Smith, Goulart, and
  Morari}]{tanaskovic2013}
Tanaskovic, M., Fagiano, L., Smith, R., Goulart, P., and Morari, M. (2013).
\newblock Adaptive model predictive control for constrained linear systems.
\newblock In \emph{2013 European Control Conference (ECC)}, 382--387.

\bibitem[{Wahlberg and M{\"a}kil{\"a}(1996)}]{wahlberg1996approximation}
Wahlberg, B. and M{\"a}kil{\"a}, P. (1996).
\newblock On approximation of stable linear dynamical systems using {L}aguerre
  and {K}autz functions.
\newblock \emph{Automatica}, 32(5), 693--708.

\end{thebibliography}

\appendix
\section{Reformulation of robust constraints} \label{sec:RobCons}
Constraints \eqref{eq:RobustCostCons} and \eqref{eq:OutputCons} contain optimization problems which are bilinear in the variables $ \phi $ and $ h $. Consider one such optimization problem, 
\begin{equation}\label{eq:primal1}
\begin{tabular}{r l}
$ \gamma = \displaystyle\max_{h} $ &$  \: [\phi(i|t)\tr] h $\\[0.5em] 
s. t.& $ A_h h \le b_h(t)$, \\
\end{tabular}
\end{equation}
where $ \gamma $ is the optimal value of the objective function and $ i \in [t,t+N+m-1] $. Then, the dual of the problem \eqref{eq:primal1} can be written as
\vspace{0.5em}
\begin{equation}\label{eq:dual1}
\begin{tabular}{r l}
$  \tilde{\gamma}= \displaystyle\min_{\theta_i}  $ &$ \: b_h(t)\tr\theta_i $\\[0.5em] 
s. t.& $ A_h\tr \theta_i = \phi(i|t), $\\
\end{tabular}
\end{equation}
where $ \theta_i \in \mathbb{R}^{p} $ is the corresponding dual variable and $ \tilde{\gamma} $ is the optimal dual objective. According to the strong duality theorem for linear programs from \cite{bertsimas1997introduction}, it holds that $ \gamma = \tilde{\gamma} $. Hence, the output constraints of the form 
\begin{equation*}
\max_{h \in H(t)}   \phi(i|t)\tr h \le \bar{y} -\eta_m \\
\end{equation*}
can be written as the set of linear constraints
\begin{align}\label{eq:dual2}
\begin{split}
\min_{\theta_i} b_h(t)\tr\theta_i &\le \bar{y} -\eta_m\\
A_h\tr \theta_i &= \phi(i|t).
\end{split}
\end{align}
The minimization problem in \eqref{eq:dual2} can be omitted, since it is on the lower side of an inequality constraint. That is, the following set of linear constraints
\begin{align}\label{eq:dual3}
\begin{split}
b_h(t)\tr\theta_i &\le \bar{y} -\eta_m\\
A_h\tr \theta_i &= \phi(i|t),
\end{split}
\end{align}
result in the same feasible dual variables $ \theta_i $ as the constraints in \eqref{eq:dual2}. Using this procedure, \eqref{eq:RobustCostCons} and \eqref{eq:OutputCons} can be converted to a set of linear equality and inequality constraints.
%The minimization was omitted from the dual problem, since it is on the lower side of the inequality and hence redundant. Extending this procedure, using matrices $ \Theta,\tilde{\Theta} \in \mathbb{R}^{p\times N+m-2} $, the output constraints \eqref{eq:OutputCons} can be expressed as
%% terminal constraint?
%\begin{equation}\label{eq:dualCons2}
%\begin{array}{r l l}
%b_h(t)\tr \Theta_i &\le \bar{y}-\eta_m, \\[0.3em]
%A_h\tr \Theta &= -\phi(i|t), \\
%b_h(t)\tr \tilde{\Theta}_i &\le \bar{y}-\eta_m,\\ [0.3em]
%A_h\tr \tilde{\Theta} &= \phi(i|t) \\[0.3em]
%&  \quad \forall i \in [t,t+N+m-2]. 
%\end{array}
%\end{equation}
%Similarly, matrices $ \Gamma,\tilde{\Gamma} \in \mathbb{R}^{p\times N} $ can be used to express the constraints 
%\eqref{eq:RobustCostCons} as
%\begin{equation}\label{eq:dualCons1}
%\begin{array}{r l l}
%y_\text{des}(i|t) + b_h(t)\tr [\Gamma]_{*i} &\le c_i, &\\[0.3em]
%A_h\tr \Gamma &= -\phi(i|t), \\ 
%-y_\text{des}(i|t) + b_h(t)\tr [\tilde{\Gamma}]_{*i} &\le c_i,\\ [0.3em]
%A_h\tr \tilde{\Gamma} &= \phi(i|t) \\[0.3em]
%&  \quad \forall i \in [t,t+N-1] 
%\end{array}
%\end{equation} 
\end{document}